\documentstyle[epsfig,kuhlen]{qcdparis}
\begin{document}

   \def\unlock{\catcode`@=11}

   \def\lock{\catcode`@=12}

   \unlock

   \def\gsim{\mathrel{\mathpalette\@versim>}}
   \def\lsim{\mathrel{\mathpalette\@versim<}}

   \def\@versim#1#2{\vcenter{\offinterlineskip
        \ialign{$\m@th#1\hfil##\hfil$\crcr#2\crcr\sim\crcr } }}

\newcommand{\proc}[2]{\mbox{$ #1 \rightarrow #2 $}}
\newcommand{\av}[1]{\mbox{$ \langle #1 \rangle $}}
\newcommand{\W}{\mbox{$W~$}}
\newcommand{\Q}{\mbox{$Q~$}}
\newcommand{\xb}{\mbox{$x~$}}  
\newcommand{\xf}{\mbox{$x_F~$}}  
\newcommand{\xg}{\mbox{$x_g~$}}
\newcommand{\y}{\mbox{$y~$}}
\newcommand{\Qsq}{\mbox{$Q^2~$}}
\newcommand{\et}{\mbox{$E_T~$}}
\newcommand{\kt}{\mbox{$k_T~$}}
\newcommand{\zz}{\mbox{$z_p~$}}
\newcommand{\ptrans}{\mbox{$p_T~$}}
\newcommand{\ejet}{\mbox{$E_{\rm jet}$}}
\newcommand{\ktjet}{\mbox{$k_{T {\rm jet}}$}}
\newcommand{\etjet}{\mbox{$E_{T {\rm jet}}$}}
\newcommand{\xjet}{\mbox{$x_{\rm jet}$}}
\newcommand{\pth}{\mbox{$p_T^h~$}}
\newcommand{\pte}{\mbox{$p_T^e~$}}
\newcommand{\ptsq}{\mbox{$p_T^{2}~$}}
\newcommand{\nch}{\mbox{$n_{\rm ch}~$}}
\newcommand{\as}{\mbox{$\alpha_s~$}}
\newcommand{\mz}{\mbox{$m_Z~$}}
\newcommand{\ycut}{\mbox{$y_{\rm cut}~$}}
\newcommand{\ymin}{\mbox{$y_{\rm min}~$}}
\newcommand{\shat}{\mbox{$\hat{s}~$}}
\newcommand{\thjet}{\mbox{$\theta_{\rm jet}~$}}
\newcommand{\gx}{\mbox{$g(x_g,Q^2)$~}}
\newcommand{\xpart}{\mbox{$x_{\rm part~}$}}
\newcommand{\mrsdm}{\mbox{${\rm MRSD}^-~$}}
\newcommand{\mrsdmp}{\mbox{${\rm MRSD}^{-'}~$}}
\newcommand{\mrsdn}{\mbox{${\rm MRSD}^0~$}}
\newcommand{\lambdams}{\mbox{$\Lambda_{\rm \bar{MS}}~$}}
\newcommand{\cm}{\mbox{\rm ~cm}}
\newcommand{\GeV}{\mbox{\rm ~GeV~}}
\newcommand{\GeVx}{\rm GeV}
\newcommand{\MeV}{\mbox{\rm ~MeV~}}
\newcommand{\GeVsq}{\mbox{${\rm ~GeV}^2~$}}
\newcommand{\nb}{\mbox{${\rm ~nb}~$}}
\newcommand{\nbinv}{\mbox{${\rm ~nb^{-1}}~$}}
\newcommand{\mm}{\mbox{$\cdot 10^{-2}$}}
\newcommand{\mmm}{\mbox{$\cdot 10^{-3}$}}
\newcommand{\mmmm}{\mbox{$\cdot 10^{-4}$}}
\newcommand{\epem}{\mbox{$e^+e^-$}}
\newcommand{\ep}{\mbox{$ep~$}}
%
%
\addtolength{\oddsidemargin}{0.3cm}
\addtolength{\evensidemargin}{0.3cm}
\addtolength{\topmargin}{2.0cm}
\addtolength{\footnotesep}{0.2cm}
\pagestyle{plain}

\thispagestyle{headings}
\title{
Hadronic Final States in Deeply Inelastic Scattering $^{\,\ast}$ }

\author{  M. Kuhlen }

\affil{Max-Planck-Institut f\"ur Physik \\
       Werner-Heisenberg-Institut \\
       F\"ohringer Ring 6 \\
       D-80805 M\"unchen \\
       Germany \\
       E-mail: kuhlen@desy.de
    \vspace{2.2cm}   }

\abstract{
Results on hadronic final states
in deeply inelastic scattering are reviewed.
They comprise jet production
and its interpretation in perturbative QCD,
signatures to distinguish conventional QCD dynamics from
possible new features of QCD at small $x$, and
measurements of inclusive charged particle production.
Theoretical developments
such as color dipole emission and instanton induced
final states are reported on.}

\resume{
Results on hadronic final states
in deeply inelastic scattering are reviewed.
They comprise jet production
and its interpretation in perturbative QCD,
signatures to distinguish conventional QCD dynamics from
possible new features of QCD at small $x$, and
measurements of inclusive charged particle production.
Theoretical developments
such as color dipole emission and instanton induced
final states are reported on.}

\twocolumn[\maketitle]
\fnm{7}{Summary talk from the working group II ``Hadronic Final States''
at the Workshop on Deep Inelastic Scattering and QCD,
Paris, April 1995}

\hyphenation{phe-no-me-no-lo-gy}

\section{Introduction}

The basic measurement
in deeply inelastic scattering (DIS)
is a measurement of the cross section \proc{ep}{eH}
in terms of the structure function $F_2$,
where $H$ stands for any hadronic system.
A wealth of information upon the partonic structure of
the proton and its dynamics have been obtained from
structure function measurements.
Measurements of the properties of the hadronic final state $H$
provide complementary
information which cannot be obtained from
inclusive structure functions.

\begin{figure}[htb]
   \centering
   \begin{picture}(1,1) \put(0.,30.){QPM} \end{picture}
   \begin{picture}(1,1) \put(120.,30.){BGF} \end{picture}
   \epsfig{file=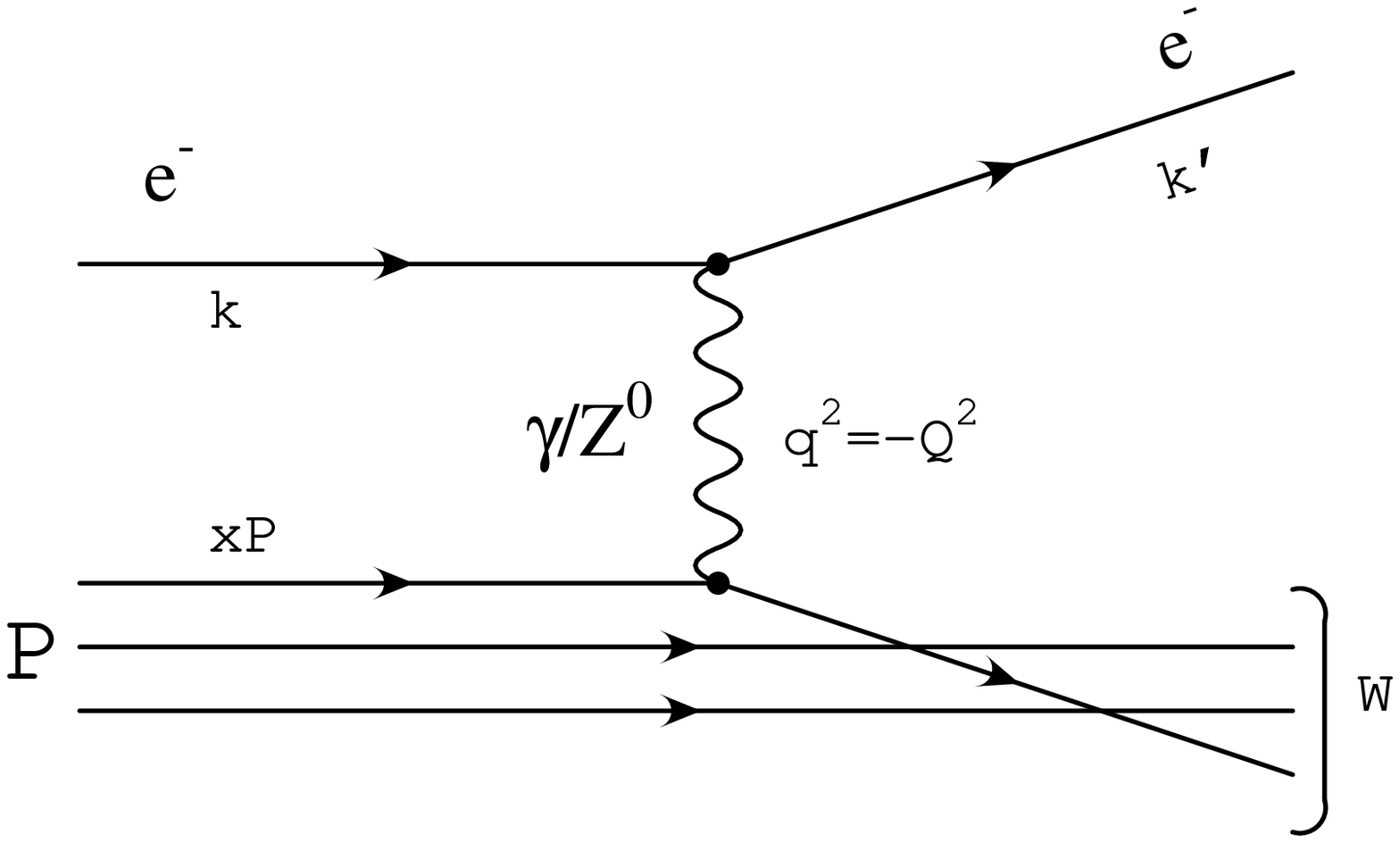,width=4cm,
    bbllx=50pt,bblly=483pt,bburx=522pt,bbury=771pt,clip=}
   \epsfig{file=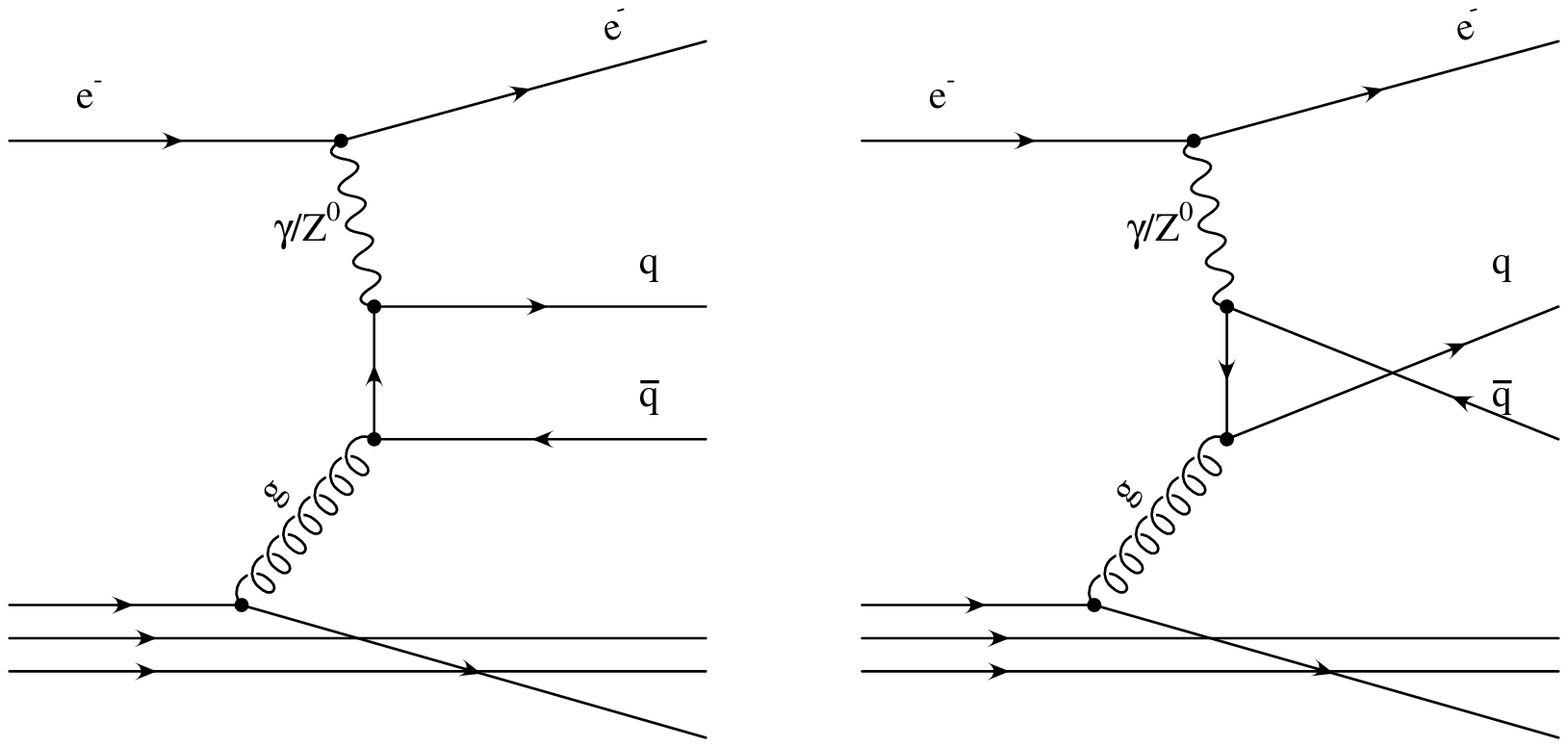,width=4cm,
    bbllx=68pt,bblly=435pt,bburx=289pt,bbury=661pt,clip=}
   \epsfig{file=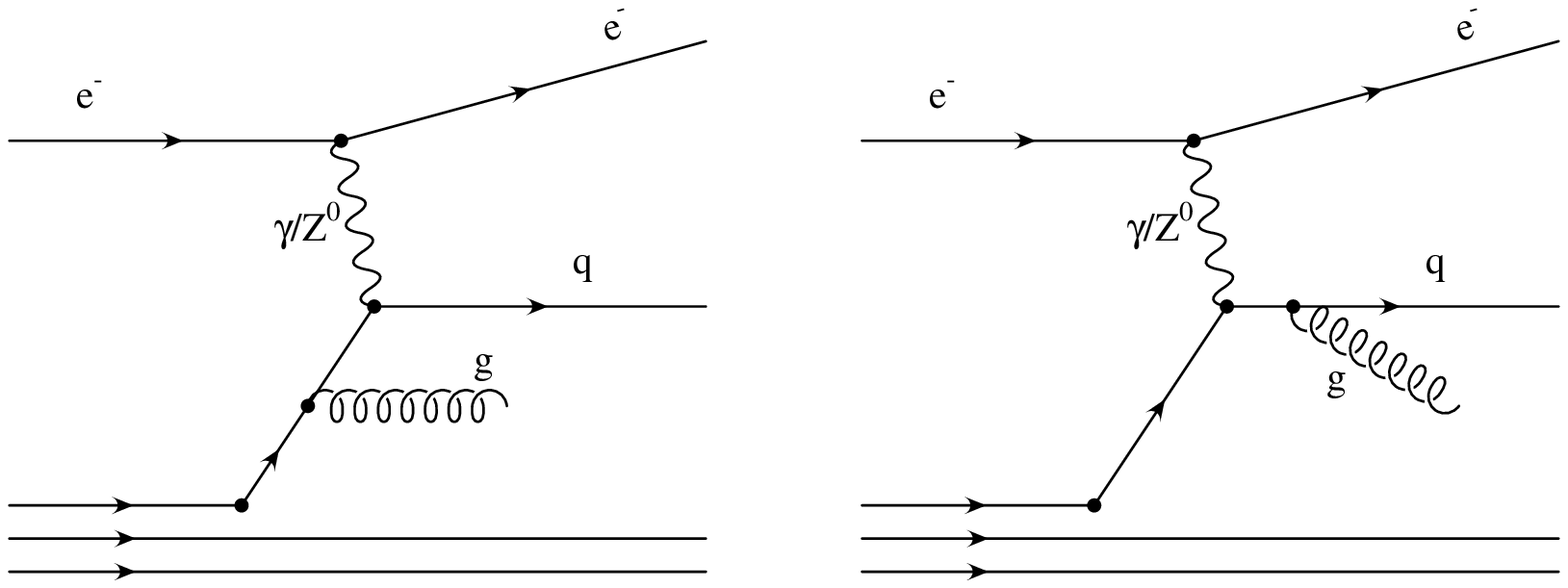,width=8 cm}
   \begin{picture}(1,1) \put(-230.,30.){QCDC} \end{picture}
   \begin{picture}(1,1) \put(-110.,30.){QCDC} \end{picture}

   \caption{\em Diagrams for DIS in $O(\as^0)$ (quark parton model - QPM)
             and in $O(\as^1)$:
             boson-gluon fusion (BGF)
             and QCD Compton (QCDC) processes.}
   \label{qcdgraphs}
\end{figure}

In the simple quark parton model (QPM) of DIS, a quark is scattered
out of the proton by the virtual boson
emitted from the scattering
lepton. QCD modifies this picture. Partons may be radiated
before and after the boson-quark vertex, and the boson may
also fuse with a gluon inside the proton by producing a
quark-antiquark pair (\fref{qcdgraphs}).
In fact, the parton which is probed
by the boson may be the end point in
a whole cascade of parton branchings.
This parton shower materializes in the hadronic final state,
allowing experimental access to the dynamics governing the
cascade.

HERA has opened a new kinematic domain to study QCD in DIS,
and most contributions in this working group were
concerned with HERA physics.
In HERA electrons of $E_e \approx 27 \GeV$ collide with
protons of $E_p=820 \GeV$, resulting in a centre of mass energy
of $\sqrt{s}\approx 300 \GeV$.
The kinematic region covered with the present data is roughly
$10^{-4}<x<10^{-1}$, $7 \GeVsq<Q^2<5000 \GeVsq$ and
$40\GeV<W<300\GeV$,
where \W\ is the invariant mass of the hadronic system, \Qsq\ the
negative 4-momentum transfer squared, and \xb the Bjorken scaling
variable, to be identified with the proton momentum fraction carried
by the scattering parton.
HERA offers the opportunity to study
the evolution of physics quantities over a large
kinematic range.
The large phase space available for hard QCD radiation,
which can be treated in perturbative QCD,
leads
to prominent jets observable in the final state.
Another area of recent interest
is the kinematic regime at small $x$ ($x\lsim 10^{-3}$)
but sizeable \Qsq,
not accessible at pre-HERA DIS experiments,
where
novel QCD dynamics are expected to play a r\^{o}le (e.g. \cite{levin}).

The two HERA detectors ZEUS and H1 \cite{detectors}
are large multipurpose,
``almost 4$\pi$'' detectors built around the beam line.
Inner tracking detectors
for charged particle detection are surrounded by a magnet and calorimetry.
Both the scattered electron serving as a tag for DIS events and the hadronic
final state are measured.
Note that a substantial part of the hadronic final state,
the proton remnant, leaves the detectors unobserved in
the beam pipe. The region close to the proton beam direction is
often referred to as the forward region.

Apart from the laboratory frame,
the hadronic centre of mass
system (CMS) and the Breit frame are used in the analyses.
The Breit frame
is defined by the condition that the virtual photon does not transfer
energy, only momentum. In the QPM picture the scattering quark would thus
just reverse its momentum of magnitude $Q/2$.
The
CMS is defined as the centre of mass system of the incoming
proton and the virtual boson, i.e. the CMS of the hadronic final
state with invariant mass $W$.
In both systems
the hemisphere defined by the virtual photon direction is referred
to as the current region, the other (containing the proton remnant)
as the target region.
The CMS current and target systems are back to back with momentum
$W/2$ each.
Longitudinal and transverse quantities are calculated w.r.t. the
boson direction.
With a longitudinal boost
from the Breit frame into the CMS, particles formerly assigned
to the target hemisphere may now end up in the current hemisphere.

Monte Carlo (MC) models based upon QCD phenomenology are used to simulate
the DIS process.
The MEPS model (Matrix Element plus Parton Shower),
an option of the LEPTO generator \cite{lepto},
incorporates the QCD matrix elements up to first order, with additional
soft emissions generated by adding leading log parton showers.
In the colour dipole model (CDM) \cite{dipole,ariadne}
radiation stems from
a chain of independently radiating dipoles formed by
the colour charges.
Both programs
use the Lund string model \cite{string} for hadronizing the
partonic final state.
Deficiencies of the Herwig parton shower model \cite{herwig}
have now been fixed
by adding matrix element corrections \cite{seymour,webber}.
This model implements an alternative
(cluster) fragmentation scheme \cite{cluster},
allowing for valuable cross
checks in the future.

\section{Jet physics}

The processes contributing to DIS up to first order in \as are shown
in figure \ref{qcdgraphs}. The QPM process results in a so-called
``1+1'' jet topology, while the QCDC and BGF processes give
``2+1'' jet events, where the ``+1'' refers to the unobserved remnant jet.
{}From a measurement of the 2+1 jet rate at large
\xb and \Qsq, where the parton densities are well known,
\as can be measured.
At small \xb and \Qsq, one can
determine the largely unknown gluon density from the
rate of 2+1 jet events,
which is then dominated by the BGF graph
(assuming \as to be known).
Complications arise from the fact that
the initial state contains strongly interacting particles,
leading to the evolution of parton showers.
Such effects need to be taken into account with
the help of MC simulations.

\subsection{The strong coupling constant \as}

Both H1 and ZEUS use the modified JADE algorithm \cite{jade}
with resolution parameter $\ycut=0.02$ to define jets in
the \as analysis.
A pseudoparticle is introduced to account for
the unobserved remnant, and then all particles $i,j$
satisfying $m_{ij}^2<\ycut \cdot W^2$ are merged into jets.
The chosen \ycut value is a compromise between statistical
precision (small \ycut), and controllable higher
order corrections (large \ycut) \cite{graudenz}.
In the H1 analysis \cite{h1alphas}
an angular cut $\thjet > 10^\circ $ (w.r.t. the proton direction)
protects against
parton showers close to the remnant.
The obtained jet rates are corrected for detector effects,
remaining parton shower contributions and hadronization with the MEPS model.
In order to extract \as from the measured jet rates, it is important
to take next to leading order (NLO) corrections into account to
reduce dependencies upon \ycut and the chosen renormalization and
factorization scales \cite{graudenz}.
Using
PROJET \cite{projet} as
NLO calculation,
the measured jet rate then yields measurements of $\as(Q^2)$
in the range
$10 \GeVsq< \Qsq < 3000 \GeVsq$, which can be seen to
run according to the QCD expectation \cite{h1alphas}.
However, below $\Qsq=100 \GeVsq$, the corrections are very model dependent
(MEPS vs. CDM). Therefore only data at $\Qsq>100 \GeVsq$
are used to extract $\as(\mz^2)=0.123 \pm 0.018$ \cite{h1alphas}.

For 2+1 jet events with $\Qsq>160\GeVsq$ and $x>0.01$,
ZEUS has measured the jet distribution in the
Lorentz invariant \zz variable \cite{zjets},
which in the centre of mass frame
of the virtual photon and the incoming parton is an angular
variable $\zz=\frac{1}{2}\cdot (1-\cos \hat{\theta}_{\rm jet})$.
Here
$\hat{\theta}_{\rm jet}$ is the angle of the
jet w.r.t. the
direction of the incoming parton.
Perturbation theory in next to
leading order (NLO) \cite{projet} is able to describe the jet
angular distribution
down to $\zz \approx 0.1$. For
$\zz<0.1$ an excess of jets is observed.
Both, the MEPS (LO matrix element + parton showers)
and ME (pure LO matrix element) simulations are similar to the
NLO calculation \cite{zjets}. The excess of jets at $\zz<0.1$
is therefore unlikely to be cured by next to NLO calculations.

\begin{figure}[t]
   \centering
   \epsfig{file=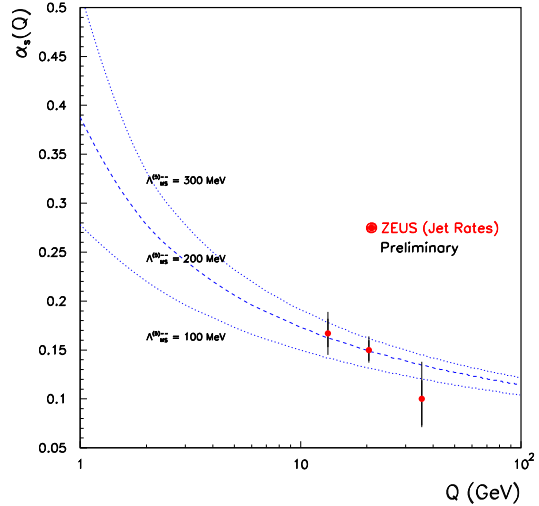,width=7cm,
    bbllx=52pt,bblly=188pt,bburx=494pt,bbury=610pt,clip=}
   \caption{\em
      Preliminary $\as (Q)$ measurements from ZEUS,
      compared to the QCD predictions
      corresponding to \lambdams =~100, 200 and 300~GeV.}
   \label{as}
\end{figure}

For the \as extraction,
a cut $\zz>0.1$ restricts the data
to a region well described by NLO perturbation theory and
QCD models \cite{grindhammer}.
The preliminary \as measurements \cite{grindhammer}
for $100 \GeVsq \lsim \Qsq \lsim 3600\GeVsq$
demonstrate the potential of HERA to study the dependence
of \as upon the renormalization scale,
and agree well with the QCD expectation (see \fref{as}).
It is expected that
already the analysis of the 1994 HERA data, once finalized,
will yield a very competitive measurement of $\as(\mz^2)$.

\subsection{The gluon density in the proton}

The 2+1 jet sample (defined with the cone algorithm in the CMS)
in the range $10\GeVsq<\Qsq<100 \GeVsq$ is used to
extract the gluon density \gx, because there the BGF
graph (\fref{qcdgraphs}) dominates
(BGF:QCDC $\approx 4:1$ \cite{grindhammer,h1gx}).
The momentum fraction \xg
which the gluon carries is calculated from the invariant mass$^2$ \shat
of the
hard subsystem forming the 2 jets via
$\xg=x(1+\hat{s}/\Qsq)\approx \hat{s}/W^2$.
Special cuts remove events affected by parton showers
\cite{grindhammer,h1gx}.
The MEPS model is used to unfold detector effects,
the QCDC contribution, QPM background and
remaining parton
shower contributions.
The MEPS model
employs a cut-off for invariant parton-parton masses
$m_{ij}^2>\ymin\cdot W^2$
to regulate
divergencies of its LO matrix element.
In order to access \xg as small as possible,
\shat is chosen as small as experimental resolution allows,
and as problems with the diverging LO matrix
element can be avoided.
It has to be ensured that the
BGF events to be analyzed
are actually generated by the model and
do not fall below that cut-off \cite{grindhammer,h1gx}.

\begin{figure}[t]
   \centering
   \begin{picture}(1,1) \put(160.,55.){prelim.} \end{picture}
   \epsfig{file=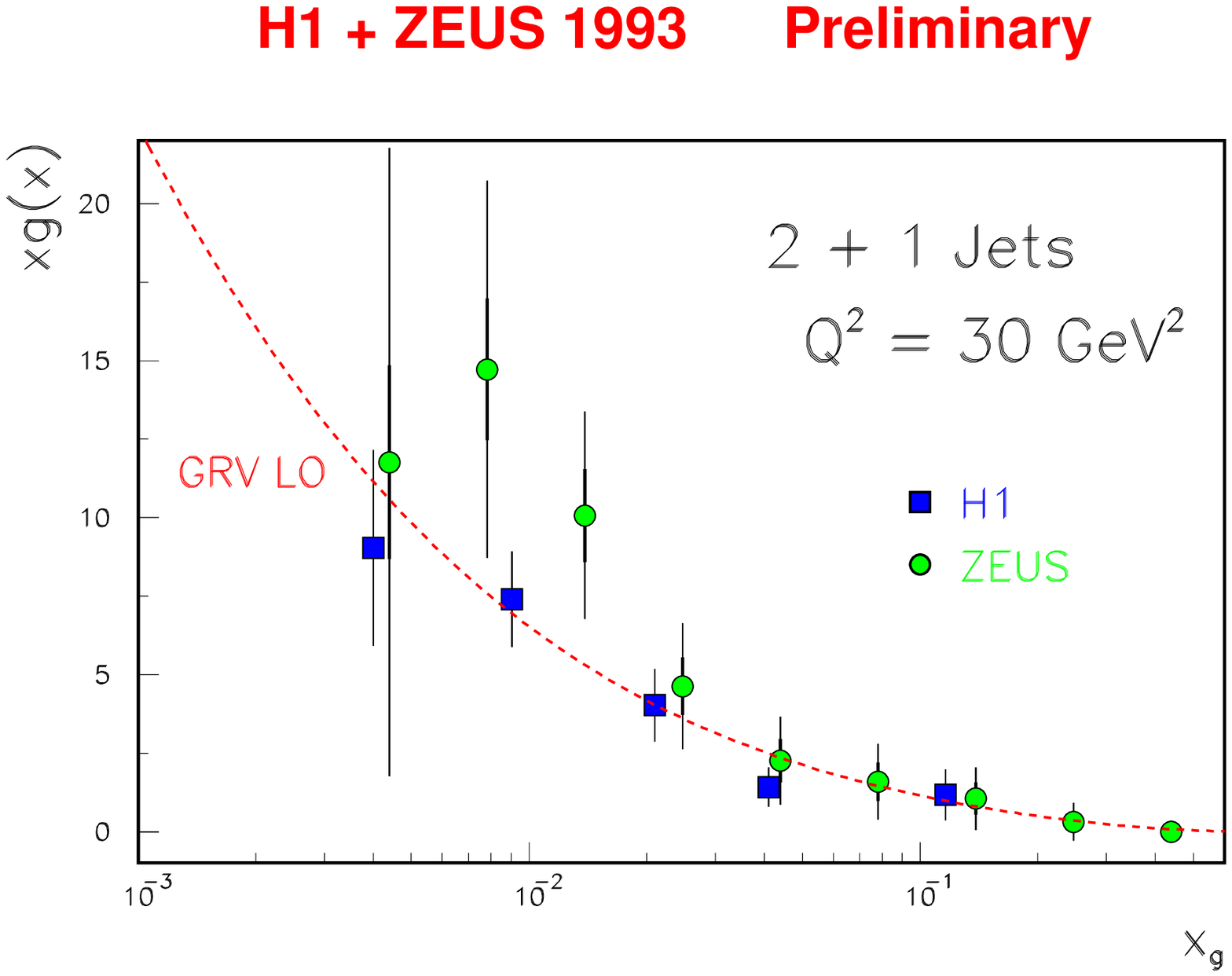,width=7cm,
    bbllx=25pt,bblly=291pt,bburx=520pt,bbury=631pt,clip=}
   \caption{\em
       The gluon density in the proton, determined in leading order (LO)
       from the rate
       of 2+1 jet events. Shown are data from H1 and ZEUS, compared
       to the LO GRV \protect\cite{grv} gluon density parametrization.}
   \label{xgluon}
\end{figure}

The H1 analysis \cite{grindhammer}
uses a fixed cut-off $\hat{s}>100 \GeVsq$ to define
BGF events, and they parametrize the MEPS cut-off such
as to follow
the limit
at which the order \as contribution
exceeds the total cross section
within a margin of $\Delta \sqrt{\hat{s}} = 2 \GeV$.
ZEUS uses the standard \ymin cut-off scheme in the MEPS model and defines
BGF events via $\hat{s}>\ymin \cdot W^2$. The parameter \ymin is then
varied between 0.0025 and 0.01
to study its influence on the result.
The H1 and ZEUS results \cite{grindhammer}
agree well with each other,
but yield different size systematic errors
(figure~\ref{xgluon}).
The ZEUS errors receive large contributions from the \ymin variation.
The
rise of the measured gluon density
towards small \xb
can be
described by a LO gluon density \cite{grv} following the
DGLAP (Dokshitzer-Gribov-Lipatov-Altarelli-Parisi)
\cite{dglap} equations.
The data are also consistent with the indirect
determination of \gx from the scaling violations of $F_2$ \cite{qcdfit},
providing a non-trivial test of QCD.
\subsection{Open Points}



Lack of understanding of parton showers close to the remnant
(model dependent corrections, failure of NLO calculations)
currently prevents
the \as analysis to make full use of the large statistics data at
$\Qsq < 100 \GeVsq$.
Though increasing HERA luminosity
will allow the \as analysis to be restricted to
higher \Qsq to reduce uncertainties,
the understanding of the forward region remains a challenge.

So far the \as measurements rely solely upon the JADE algorithm,
being the only algorithm for which NLO jet cross sections are
calculated \cite{nlo}.
NLO calculations
for other algorithms, such as the cone \cite{cone}
or the theoretically preferred $k_T$ \cite{kt} algorithm
are desirable.
Such a program,
which would also be able to calculate
event shape variables
like energy-energy correlations, Thrust, etc.,
is being worked upon by D. Graudenz, but results
cannot be expected in a short term.
Theoretical uncertainties could also be reduced by resumming
higher order corrections.

The validity of corrections from hadronic to partonic
final states, defined either in LO or NLO,
need to be checked with models based upon different
parton shower and hadronization schemes.
Unfortunately, a MC generator incorporating the QCD matrix
elements beyond LO is missing.

The gluon density has so far been determined in LO. A method
allowing a measurement in NLO is presently under study \cite{graudenz}.

How can \as be determined consistently, considering it is input
for the evolution of parton densities which are used
in the analysis \cite{vogt}?


\section{Novel QCD dynamics}

\begin{figure}[t]
   \centering
   \vspace{-0.2cm}
   \epsfig{file=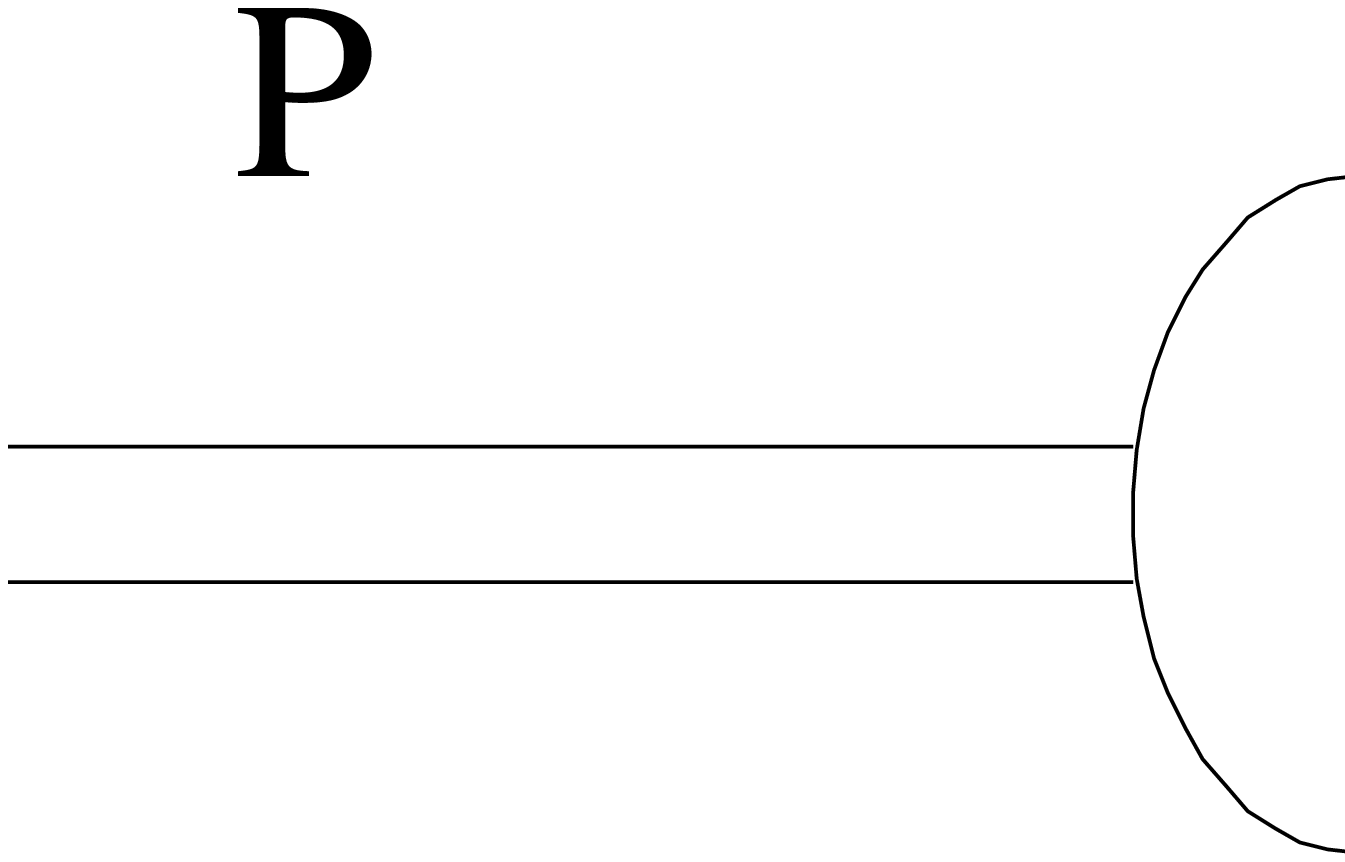,width=2cm}
   \caption{\em
      Parton evolution in the ladder approximation.
      The selection of forward jets in DIS events is illustrated.}
   \label{cascade}
\end{figure}

The observed strong rise of the structure function $F_2$ towards small
$x$ \cite{f2}
has caused much debate on whether the QCD evolution of the
parton densities can still be described by the conventional
DGLAP \cite{dglap} equations, or whether the HERA data extend into a new regime
at small $x$ where the dynamics is governed by the BFKL
(Balitsky-Fadin-Kuraev-Lipatov) \cite{bfkl} equation.
It would be extremely interesting to test QCD in such a new regime.
While the rise is consistent with the expectation from BFKL dynamics,
it can however also be described by a DGLAP evolution \cite{akms}.
At lowest order the BFKL and DGLAP equations resum the leading
logarithmic $(\as \ln 1/x)^n$ or $(\as \ln (\Qsq/ Q_0^2))^n$ contributions
respectively. In this approximation the leading diagrams are of the
ladder type (\fref{cascade}). The leading log
DGLAP ansatz corresponds to a strong
ordering of the transverse momenta \kt (w.r.t. the proton beam)
in the parton cascade
($Q_0^2 \ll \kt_1^2 \ll ... \kt_j^2 \ll ... \Qsq$),
while
there is no such ordering in the BFKL ansatz
($\kt_j^2 \approx \kt_{j+1}^2$)
\cite{ordering}.
Measurements on the hadronic final state emerging from the cascade
therefore offer another handle to search for signatures of the BFKL
behaviour.
They are compared to analytical calculations as
well as to the QCD models MEPS and CDM.
The CDM description
of gluon emission is similar to that of the BFKL evolution,
because the gluons emitted by the dipoles
do not obey strong ordering in \kt~\cite{bfklcdm}.
The MEPS model with its leading log parton shower is
based upon DGLAP dynamics, and the emitted partons
are thus ordered in \kt.

\subsection{Transverse Energy Production}

As a consequence of the strong \kt ordering the DGLAP
evolution is expected to produce less transverse energy \et in a
region between the current region and the proton remnant
than the BFKL evolution \cite{durham}.
H1 and ZEUS have measured the flow of transverse
energy in the laboratory frame
as a function of pseudorapidity $\eta = - \ln \tan (\theta/2)$,
where $\theta$ is the angle of the energy deposition w.r.t the
proton beam axis \cite{h1flow2,haas,h1flow3}.
The measurements are made
for varying ranges in \xb
($2 \mmmm < \av{\xb} < 5 \mmm$)
and \Qsq
($7\GeVsq < \av{\Qsq} < 30 \GeVsq$
and agree well between the experiments
\cite{haas}.

The \et flows
for large \xb and \Qsq are reasonably well described by MEPS and CDM.
For smaller \xb and \Qsq
both models predict a more pronounced enhancement
in the current fragmentation region than is seen in the data.
Between the current
system and the proton remnant
(the lab. forward region),
the data are reasonably well described
by the CDM, while the MEPS model produces too little \et \cite{h1flow2,haas}.
This intermediate region is expanded in \fref{et93},
because there perturbative calculations,
based either on DGLAP or on BFKL dynamics, are available \cite{durham}.
The BFKL calculation comes out
close to the data, while the DGLAP calculation predicts
much less \et.
However, the non-perturbative hadronization phase is missing
in these calculations.

\ffig{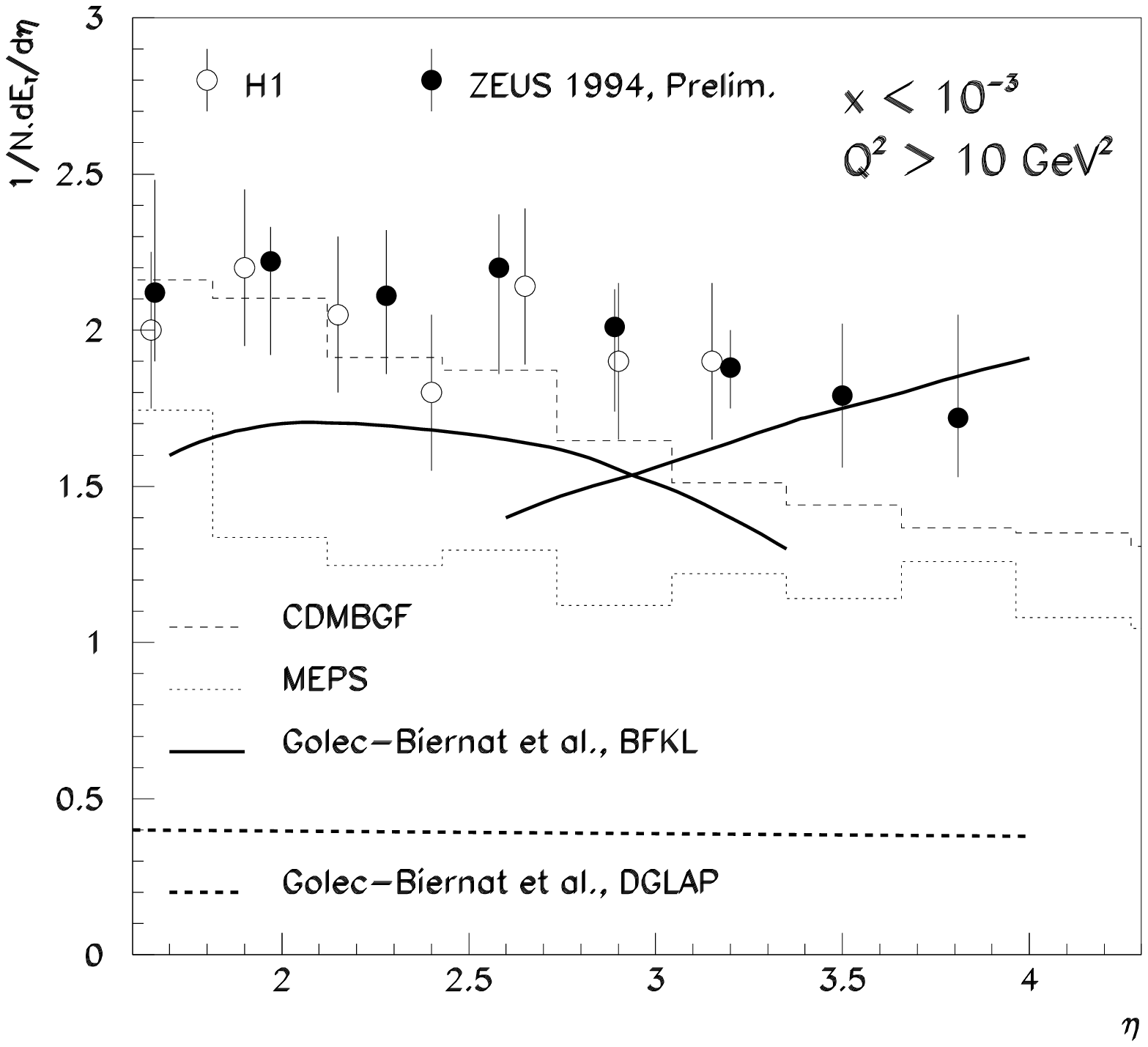}{60mm}
{\em Transverse energy flow in the forward region at H1
\protect\cite{h1flow2}
and ZEUS \protect\cite{haas} for $x<10^{-3}$.
The proton direction is to the right. The calorimeter acceptances
end at $\eta$ around 3.5.
The data are compared to the CDM (here labelled CDMBGF)
and MEPS models and to partonic
calculations based upon the DGLAP and BFKL equations
\protect\cite{durham}.}
{et93}

H1 has determined the average \et,
measured centrally
in the CMS
as a function of \xb and \Qsq (\fref{etx}).
They find an increase of \av{\et} with decreasing \xb,
which is a characteristic BFKL prediction \cite{durham}.
The data are
in agreement with the BFKL calculation \cite{sutton}, if one assumes
an \et contribution from hadronization of about 0.4 GeV per unit rapidity
(independent of \xb). That estimate is taken from
the CDM, which agrees with the BFKL calculation at the parton level.

\ffig{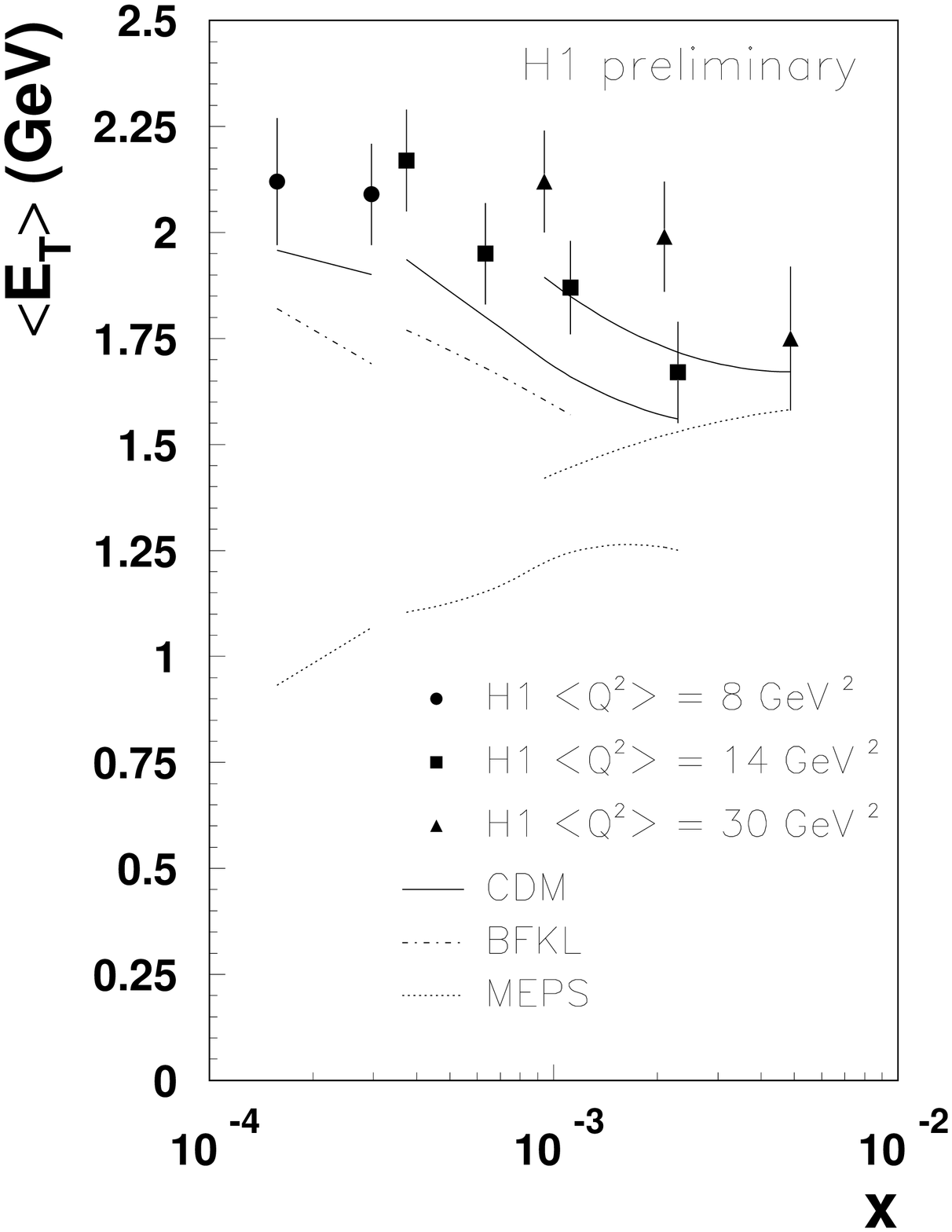}{60mm}
{\em Transverse energy  \av{\et} per unit of pseudorapidity $\eta^\ast$
as a function of \xb for three different
values of \Qsq,
measured centrally at $\eta^\ast = 0$
in the CMS (corresponding to the lab. forward region).
The data are compared to the CDM and MEPS models including
hadronization, and to the BFKL calculation (no hadronization).}
{etx}

The apparent failure of the MEPS model
has caused many questions
about its ingredients: the way the parton shower is ``matched'' to
the matrix element, the colour connection between the current and the
remnant system and its effect upon hadronization, and the remnant
fragmentation itself which is little tested.
It seems that re-arranging colour configurations can produce
enough \et through hadronization to compensate the \et deficit
in the DGLAP cascade of the MEPS model \cite{ingelman}.
A MEPS version thus modified should be available soon for
detailed testing.
The flexibility in the hadronization modelling presently
precludes unambiguous tests of the DGLAP evolution through
\et measurements.
For
the same reasons the intriguing success of the CDM without \kt ordering
may be fortuitous.
A MC model invoking explicitly the BFKL evolution,
currently being developed by K. Golec-Biernat et al.,
would help interpreting the data.
In any case, the \et data provide important input for QCD
phenomenology.

\subsection{Forward Jets}

At present strong conclusions upon the validity of the BFKL or DGLAP
parton evolutions at small $x$ from the \et measurements are hampered
by the uncertainties about hadronization.
Jet production should be less affected by hadronization. A signature
for BFKL dynamics proposed by \cite{mueller} is the production of
``forward jets'' with $\xjet=\ejet/E_p$, the ratio of jet energy and
proton beam energy,
as large as possible, and with
transverse momentum \ktjet ~close to \Q in order to reduce the phase
space for the \kt ordered DGLAP evolution (see \fref{cascade}).
An enhanced rate of events with such jets is thus expected in the BFKL
scheme \cite{mueller,dhotref}.
The experimental difficulty is to detect these
``forward'' jets which are close to the beam hole in proton
direction.

\begin{figure}[t]
   \centering
   \begin{picture}(1,1) \put(40.,40.){preliminary} \end{picture}
   \epsfig{file=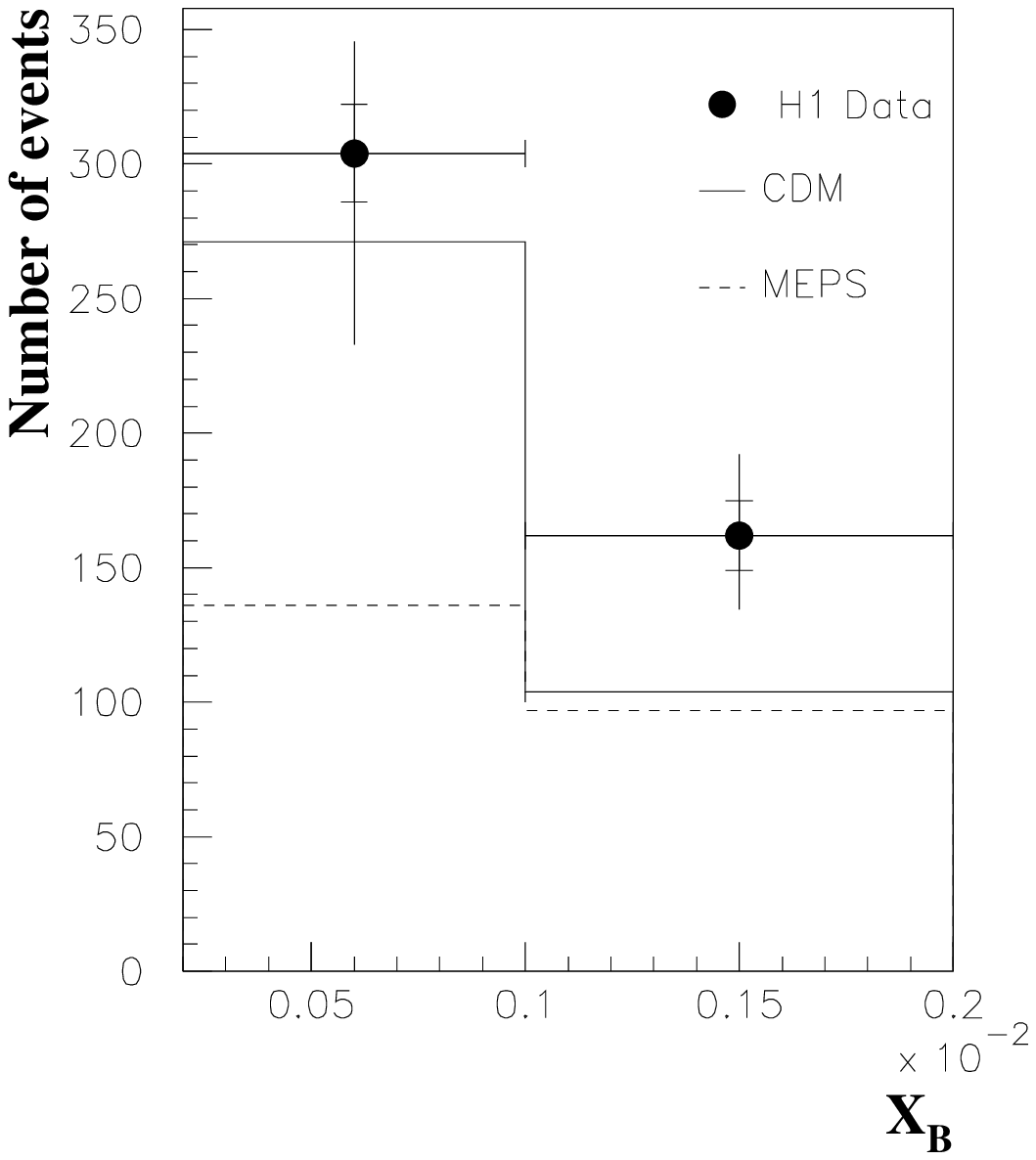,width=5cm}
   \caption{\em
     The rate of forward jets
     (selected with $\xjet>0.025$,
     $0.5 < \ktjet^2/\Qsq <4$ and $\ktjet>5 \GeV$) in the kinematic
     range $2 \mmmm < \xb < 2 \mmm$ and $\Qsq \approx 20 \GeVsq$.
     The measurement is compared to the CDM and MEPS models.}
   \label{fwdjets}
\end{figure}

The rate of forward jets measured by H1 \cite{deroeck,h1flow3}
(\fref{fwdjets})
is larger at low \xb
than at high $x$.
This is expected from BFKL
calculations, in contrast
to calculations without BFKL ladder \cite{dhotref,delduca}.
The behaviour of the data is better represented by the CDM
than by the MEPS model.
However, neither of them describe the
energy spectrum of the observed jets correctly, and the model predictions
for the jet rates are thus cut dependent \cite{deroeck,h1flow3}.
The analysis of a larger
statistics sample should allow more firm conclusions.

\begin{figure}[t]
   \centering
   \vspace{-3.8cm}
   \begin{picture}(1,1) \put(80.,90.){ZEUS preliminary} \end{picture}
   \epsfig{file=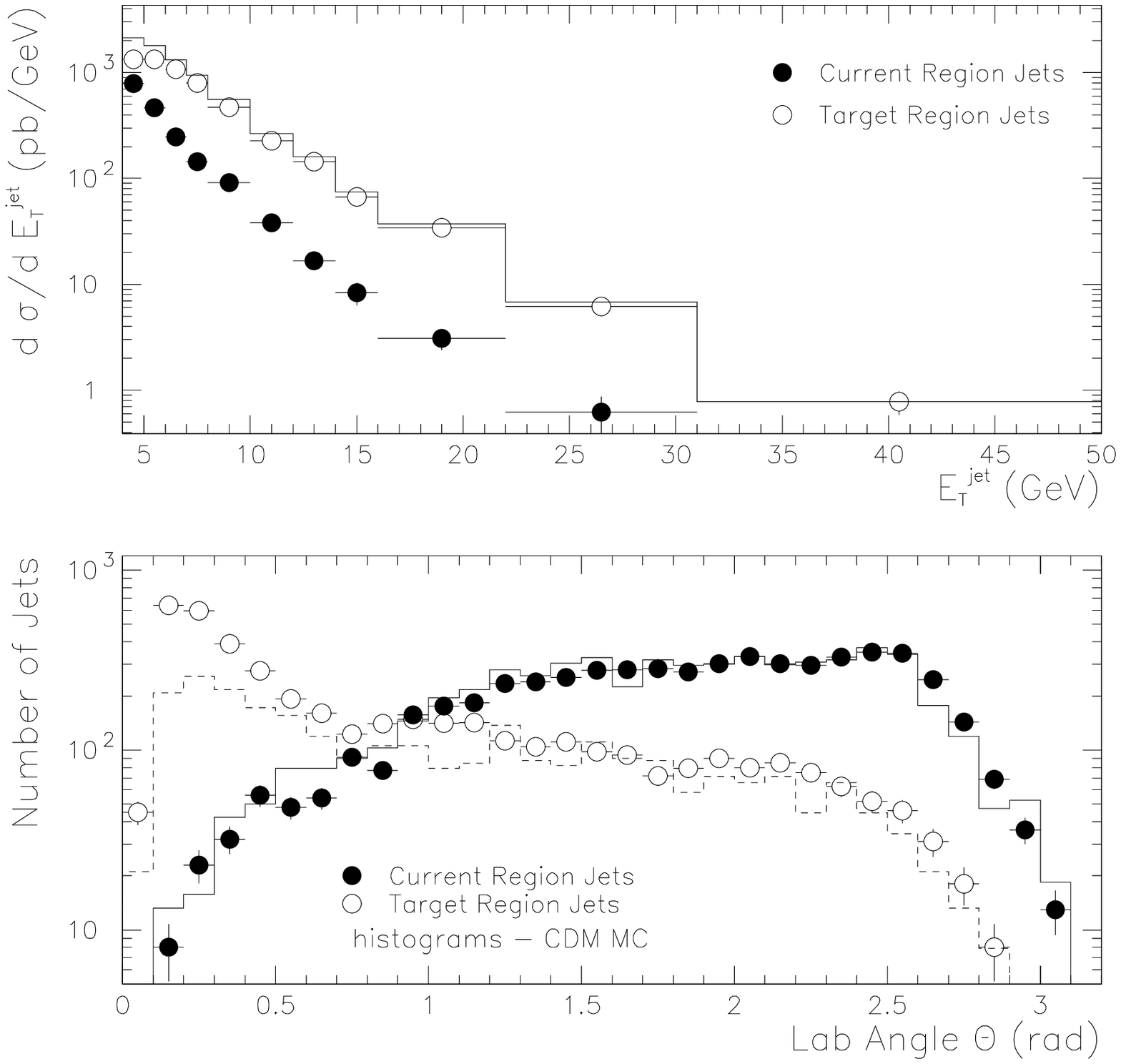,width=8cm,
    bbllx=22pt,bblly=165pt,bburx=522pt,bbury=645pt,clip=}
   \caption{\em
     The laboratory angular distribution of jets detected either in
     the Breit current or target hemisphere.}
   \label{thjet}
\end{figure}

ZEUS has measured an inclusive jet cross section
${\rm d}\sigma / {\rm d} \etjet$ in the Breit frame \cite{deroeck}.
Many more jets are found in the target region with a harder \etjet~
spectrum than in the current region, reflecting the differences
in phase space in the two regions.
The current region data are reasonably well described by the
CDM and MEPS models.
In the target region however there is a substantial excess of
jets over the model predictions,
which can be linked with an excess of $2+1$ jet events \cite{deroeck}.
In the laboratory frame
this excess is located in the forward region at angles
$\theta_{\rm jet}<20^\circ$ (\fref{thjet}).

The data on jet production in the forward region
(lab. frame), or the target region (Breit frame)
certainly pose a challenge to theory.
So far cross sections are calculated
\cite{dhotref,delduca} only for partons, while
experiments measure
hadron jets.
This gap has to be bridged from both sides to
allow a strictly valid comparison.
%

\subsection{Jet correlations}

Apart from calculations of forward jet rates, Del Duca \cite{delduca}
discussed
angular correlations for forward jets.
If a BFKL ladder is inserted in between the electron-photon vertex
and the forward jet,
the angular correlation between the forward jet and the electron
imposed by momentum conservation is relaxed.
Such angular decorrelation could be another footprint of BFKL dynamics.
The fact that 4\% of the H1 forward jet events contain a second
forward jet \cite{deroeck}
opens another route of investigation, namely correlations
between such jets.
If these jets can be identified with gluons emitted from
the ladder, it would be possible to check the parton ordering
directly.

\subsection{Dipole emission}

An interesting ansatz
to calculate final state observables was
presented by R. Peschanski \cite{peschanski}.
The starting point is
onium-onium scattering \cite{onium}
with onium wave functions which can be
derived from QCD. Such a reaction is analogous to an interaction
of the current system with the remnant system in DIS.
Radiation is treated in the dipole picture, leading to a copious
production of dipoles in the central rapidity region of the interaction.
Once such an ansatz yields
quantitative predictions,
it could be tested in DIS, e.g.
with \et flow measurements.

Bo Andersson \cite{andersson}
discussed DIS final states in terms of a chain
of radiating colour dipoles, and its connection with
the Ciafaloni-Catani-Fiorini-Marchesini ansatz \cite{ccfm}.
In principle this model could provide
a complete picture of the hadronic final state in DIS.
The implementation in the Ariadne
\cite{ariadne} MC generator is in progress
to allow detailed predictions.

\subsection{QCD Instantons}

The
standard model contains processes which cannot be
described by perturbation theory, and which violate
classical conservation laws like baryon and lepton number
in the case of the electroweak sector and chirality for
the strong interaction \cite{thooft}.
Such anomalous processes are induced by instantons \cite{belavin}.
At HERA, QCD instantons may lead to observable effects
in the hadronic final state in DIS \cite{ringwald,schrempp},
which were discussed by F. Schrempp.
The instanton should decay isotropically into a high
multiplicity state of gluons and all quark flavours
simultaneously which are kinematically allowed.
A MC program to simulate instanton events
has become available \cite{gibbs}.
Due to the isotropic decay, one expects a densely populated
region in rapidity, other than the current jet, which is
isotropic in azimuth. The presence of strangeness and charm
could provide an additional signature.
However remote the a priori chances to see such signals may appear,
here is a chance for a major discovery at HERA!

\section{Charged Particle Spectra}

The H1 and ZEUS measurements of inclusive charged particle spectra \cite{pavel}
are performed either in the Breit
frame or in the CMS.
In the Breit frame in- and outgoing quark have equal but opposite sign
momenta $Q/2$ (QPM picture),
and in \epem annihilation the outgoing quark and antiquark
have equal but opposite momenta $\sqrt{s}/2=Q/2$.
Due to this similarity it is interesting to compare
particle spectra in the Breit current hemisphere in DIS with
\epem data. DIS experiments have the advantage over \epem experiments
that they cover a large span in \Q, presently from 3 \GeV to
50 \GeV, in a single experiment. The current mean charged multiplicity
at HERA rises $\sim \ln Q$ within errors,
and agrees with \epem data (divided by 2) where they overlap
\cite{breit,pavel}.

Colour coherence should lead to a suppression of soft gluon
emission. The HERA data \cite{breit,pavel}
on the scaled charged particle momentum
distribution $\ln 1/x_p$ with $x_p=2\cdot p/Q$ exhibit the
expected hump
backed plateau \cite{basics},
the evolution of which with \Q is in agreement with
the assumption of colour coherence.
However, like in \epem annihilation,
this behaviour can also be mimicked
through the Lund string fragmentation
\cite{pavel}.

The scaled momentum spectrum of \xf in the CMS, where the
particle longitudinal momenta $p_z$
are divided by the maximal possible momentum,
$\xf=2 \cdot p_z/W$, are shown in \fref{xf} for the current region
(the target region is not observed).
Comparing HERA data at
$W \approx 120 \GeV$ \cite{h1flow2,pavel} with fixed target
data at \W = 14 and 18 \GeV \cite{emc,e665},
significant scaling violations
are observed, in agreement with QCD expectations: the large
value of \W at HERA results in a large phase space for QCD radiation,
softening the \xf spectrum w.r.t. data at lower $W$.
It can be expected that such data will
be used to extract \as in the future.
\ffig{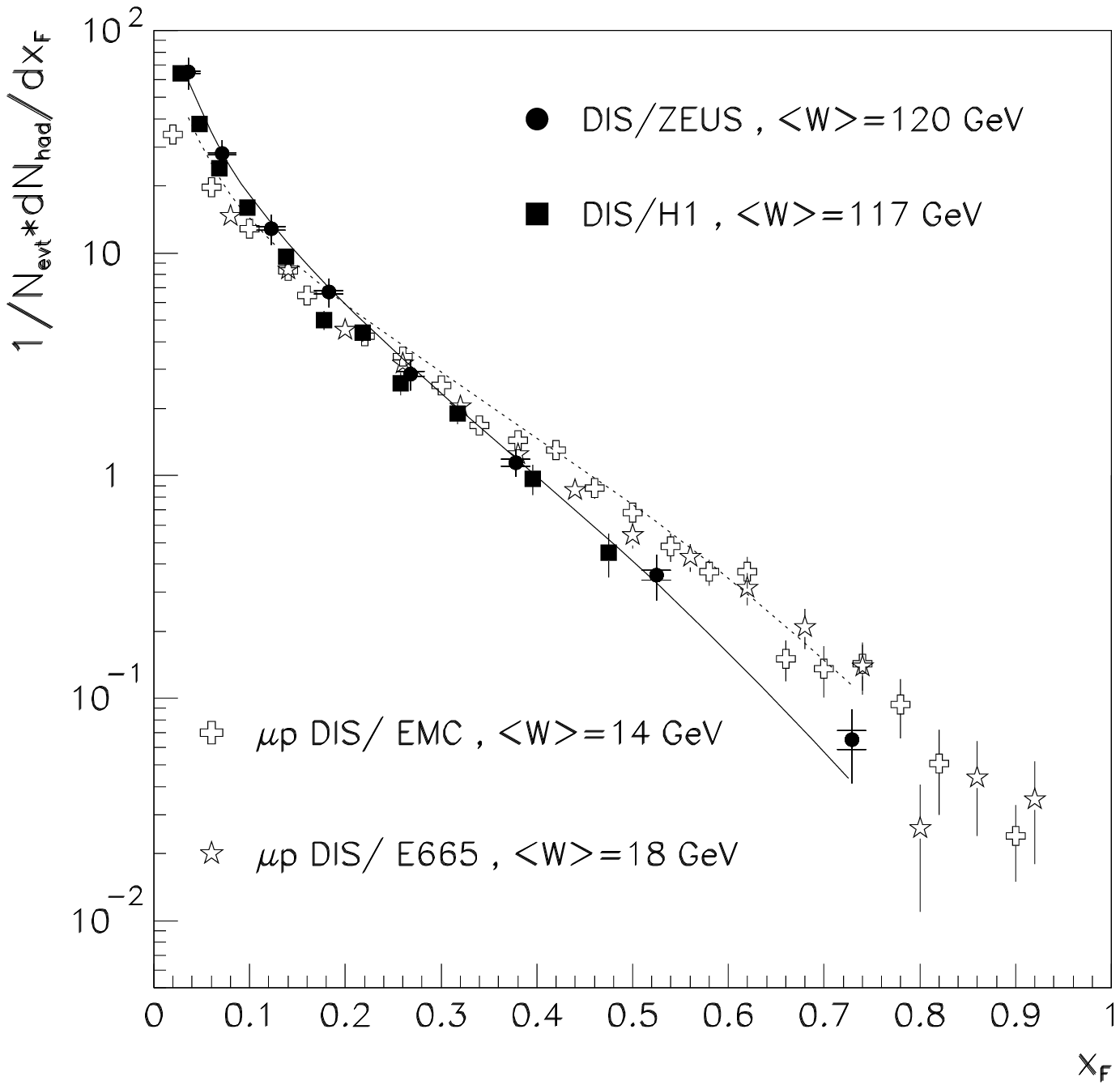}{80mm}
{\em The \xf spectra measured at HERA
compared with the QPM (dotted line) without QCD radiation,
the MEPS model (full line), and
with fixed target DIS data
at lower $W$.}
{xf}

\begin{figure}[t]
   \centering
   \epsfig{file=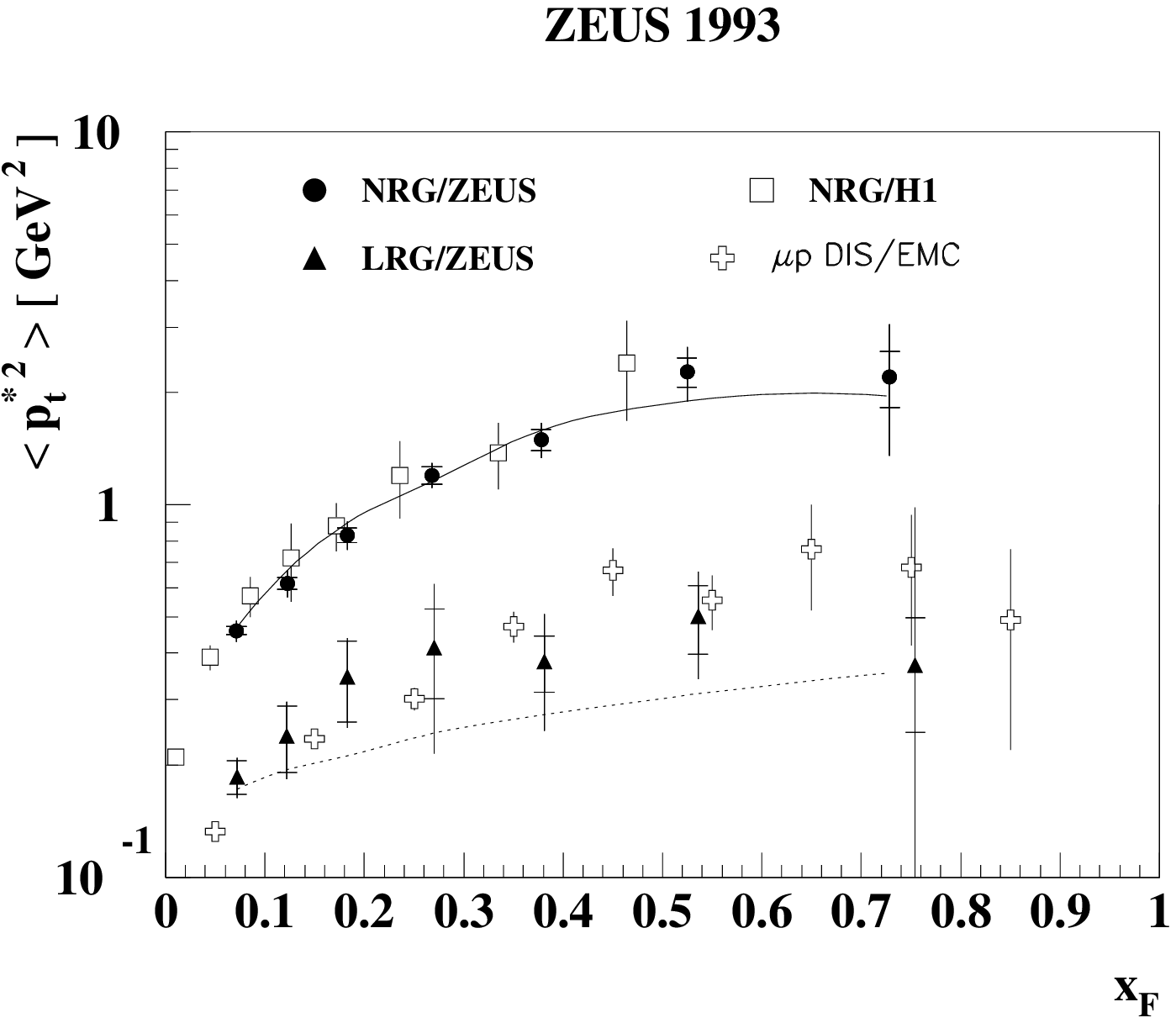,width=68mm}
   \caption{\em
      The seagull plot. Shown are the mean transverse momenta
      squared $\av{p_T^2}$
      as a function of \xf in the CMS for HERA data with and without
      a rapidity gap (LRG/NRG)
      compared
      to the QPM prediction (dotted line) and the MEPS model (full line),
      and to EMC
      data at lower $W$.}
   \label{sg}
\end{figure}

The effect of QCD radiation is clearly seen in the
``seagull plot'' (\fref{sg}), where the mean transverse momenta \ptsq
squared of the particles is plotted as a function of \xf.
As a consequence of increased
QCD radiation, much larger \ptsq are observed
at HERA \cite{h1flow2,pavel} than at EMC \cite{emc} at smaller $W$,
again in agreement with QCD expectation.
ZEUS has also compared DIS events with and without a large
rapidity gap \cite{gap} in this respect \cite{pavel}.
Much smaller $p_T^2$
than in normal DIS events are
observed in events with a large rapidity gap, thought to
stem from diffractive processes
and accounting for approximately
10\% of the total sample
\cite{gap}.
This indicates that the scale
governing radiation
is much smaller than \W for rapidity gap events.

\section{Conclusion}

Two complementary approaches to the HERA data can be
distinguished.
In one approach, one tries to identify a region which is
``well understood'', meaning that the observation agrees
with the theory and the models. Under this condition, the
data can be interpreted in the framework of the theory,
and physical quantities which are defined within the theory
can be extracted. The measurements of \as and \gx fall into
this category.
However, we have also seen data which are not yet understood
theoretically, namely hadron and jet production
in the forward region.
Such
data currently pose a challenge to the theory, and experimentalists
should make every effort to provide theory with solid data
to work with.

\begin{center}
{\large\bf Acknowledgements}
\end{center}
I would like to thank my fellow conveners, A. Doyle and G. Ingelman,
for the pleasant cooperation, the organizers of the workshop for their
efficient support and the participants of the session
for their contributions  and inspiring discussions in the working group.

%

%

\end{document}